\documentclass[sigconf]{acmart}

\usepackage{balance}
\usepackage{tabularx}
\usepackage{framed,enumitem} 
\usepackage{graphicx}
\def\BibTeX{{\rm B\kern-.05em{\sc i\kern-.025em b}\kern-.08emT\kern-.1667em\lower.7ex\hbox{E}\kern-.125emX}}
    
\copyrightyear{2020}
\acmYear{2020}
\setcopyright{rightsretained}
\copyrightyear{2020}
\acmYear{2020}
\setcopyright{rightsretained}
\acmConference[AIES '20]{Proceedings of the 2020 AAAI/ACM Conference on AI, Ethics, and Society}{February 7--8, 2020}{New York, NY, USA}
\acmBooktitle{Proceedings of the 2020 AAAI/ACM Conference on AI, Ethics, and Society (AIES '20), February 7--8, 2020, New York, NY, USA}
\acmDOI{10.1145/3375627.3375803}
\acmISBN{978-1-4503-7110-0/20/02}

\begin{document}

\fancyhead{}

\title{Beyond Near- and Long-Term: Towards a Clearer Account of Research Priorities in AI Ethics and Society}

\author{Carina Prunkl}
\authornote{Both authors contributed equally to this paper.}
\affiliation{
  \institution{Future of Humanity Institute\\University of Oxford}
  \streetaddress{Littlegate House, 16--17 St Ebbe's Street}
  \postcode{OX1 1PT}
}
\email{carina.prunkl@philosophy.ox.ac.uk}
\orcid{1234-5678-9012}

\author{Jess Whittlestone}
\authornotemark[1]
\affiliation{
  \institution{Leverhulme Centre for the Future of Intelligence\\ University of Cambridge}
}
\email{jlw84@cam.ac.uk}

%
\renewcommand{\shortauthors}{Prunkl and Whittlestone}


\begin{abstract}
One way of carving up the broad `AI ethics and society' research space that has emerged in recent years is to distinguish between `near-term' and `long-term' research. While such ways of breaking down the research space can be useful, we put forward several concerns about the near/long-term distinction gaining too much prominence in how research questions and priorities are framed. We highlight some ambiguities and inconsistencies in how the distinction is used, and argue that while there are differing priorities within this broad research community, these differences are not well-captured by the near/long-term distinction. We unpack the near/long-term distinction into four different dimensions, and propose some ways that researchers can communicate more clearly about their work and priorities using these dimensions. We suggest that moving towards a more nuanced conversation about research priorities can help establish new opportunities for collaboration, aid the development of more consistent and coherent research agendas, and enable identification of previously neglected research areas.
\end{abstract}



\maketitle
\section{Introduction}

There is a growing community of researchers focused on ensuring that advances in AI are safe and beneficial for humanity (which we call the AI ethics and society community, or AI E\&S for short.) The AI E\&S research space is very broad, encompassing many different questions: from how AI is likely to impact society in practice to what ethical issues that might raise, and what technical and governance solutions might be needed. These questions are approached from many different perspectives and disciplines: philosophy, computer science, political science, sociology, law, economics, and international relations to name just a few. This research space is fast growing, with new initiatives, workshops, and research centres emerging every year.

Finding ways to break down this broad and evolving research space is important, as it helps researchers to clearly formulate and communicate problems and research agendas, collaborate effectively with others, and to identify gaps in current research. Natural ways to break down the AI E\&S research space include by themes or sectors (such as the impact of AI on medicine, the military, or issues of social justice) or in relation to pre-existing disciplines (distinguishing research on ethics from technical approaches from governance and policy, for example).

Due to the cross-cutting nature of AI's impacts, and the need for deep interdisciplinary collaboration in this space, it may be useful to have ways of carving up the research space that relate to more fundamental priorities and assumptions rather than being tied to specific domains or disciplines. One such way of carving up the research space which has emerged in recent years is to distinguish between those who focus on `near-term' and `long-term' issues \cite{brundage_guide_2017,cave_bridging_2019,krakovna_is_2018,parson_artificial_2019}.

However, the distinction between `near-term' and `long-term' is being over-emphasised as a way of characterising differences in the research space. On closer inspection, the distinction is poorly defined, and is used differently across people and contexts: sometimes to refer to issues on different timescales; sometimes to issues related to different types of AI capabilities, and other times masks deeper normative and empirical disagreements. Highlighting inconsistencies in how the near/long-term distinction is used, we argue that it fails to adequately capture the complexity and nuance of different approaches and priorities in the AI E\&S research space, and risks causing confusion and fuelling conflict.

By unpacking the distinction between near- and long-term into four different dimensions, we propose some ways to more clearly conceptualise different priorities and disagreements within the AI E\&S research space. First, we distinguish between research which focuses on near/long-term capabilities as opposed to near/long-term impacts, and argue that in both cases near/long-term is a spectrum rather than a binary distinction. We show how this way of thinking about different approaches could both improve collaboration and communication between researchers as well as helping to identify important areas of research which are currently neglected. Second, we analyse some of the beliefs and disagreements which seem to underpin differing research priorities, and suggest that clarifying these beliefs could aid mutual understanding between different researchers and groups.

\section{The near/long-term distinction in practice}
The distinction between near- and long-term in AI E\&S is used both to distinguish different types of \textit{issues} one might be concerned about, as well as to highlight a divide or disagreement between two research \textit{communities}. In this section, we briefly review how the distinction is made in the literature.

\subsection{`Near-term' issues}

As the phrase `near-term' suggests, those who have written about the distinction tend to characterise near-term issues as those issues that society is already facing or likely to face very soon: Brundage \cite{brundage_guide_2017} defines near-term issues as those society is ``grappling with today'' and Cave and \'{O}h\'{E}igeartaigh \cite{cave_bridging_2019} talk in terms of ``immediate or imminent challenges'' (p.5). Examples include concerns about data privacy \cite{tucker2018privacy,zuboff2019age}, algorithmic bias \cite{hajian2016algorithmic,koene2017algorithmic}, self-driving car accidents \cite{goodall2016can,bonnefon2016social}, and ethical issues associated with autonomous weapons \cite{asaro2012banning,anderson2013law}. It is worth noting two things about these examples. First, they tend to be relatively concrete and specific, and an important feature of `near-term' issues is the fact that they are already fairly well-understood. Second, these examples tend to be directly related to recent progress in machine learning which have enabled increasing real-world applications of narrow and specialised AI systems: for example in medical diagnosis and predictive policing \cite{buch_artificial_2018,richardson_dirty_2019}. 

\subsection{`Long-term' issues}

`Long-term' issues are often characterised as those arising from advanced AI systems, or simply as issues that will arise far into the future. Brundage \cite{brundage_guide_2017} defines long-term issues as those that ``either only arise at all or arise to a much greater extent when AI is much more advanced than it is today'', while Cave and \'{O}h\'{H}igeartaigh \cite{cave_bridging_2019} point to ``longer-term concerns and opportunities that are less certain'' (p.5). Examples of long-term issues often centre around the implications of very advanced future AI with broad capabilities (`artificial general intelligence', or AGI), which achieve human or superhuman intelligence (`human-level AI' or `superintelligence'), or are in some other way radically transformative \cite{karnofsky_potential_2016}. For example, the White House Report on Preparing for the Future of Artificial Intelligence talks about ``long-term concerns of AGI'', implicitly equating the two \cite{whitehouse_2016}. While thinking about `long-term' concerns of AI may have begun with concerns about superintelligence and related notions \cite{bostrom_superintelligence_2014}, the notion has since broadened to include challenges that are likely to have long-term consequences, as opposed to merely arising far into the future. Examples include the effects of advanced, `transformative AI' on  international security, race dynamics and power relations \cite{karnofsky_potential_2016,dafoe_ai_2018,gruetzemacher_2019}.

\subsection{The divide between `near-term' and `long-term' communities}

Baum \cite{baum2018} not only distinguishes between near and long-term issues, but also points to a divide between two different communities: `presentists', who claim that ``attention should go to existing and near-term AI'', and `futurists', who argue for focusing on ``the potential for radically transformative AI''. Cave and \'{O}h\'{E}igeartaigh \cite{cave_bridging_2019} also point to ``two seemingly separate communities of researchers and technologists'' focused on near and long-term issues.

Baum suggests these two groups disagree quite fundamentally about what issues are most important to work on. This has provoked hostility between the two sides: the `near-term' camp suggesting that long-term concerns are overblown and a distraction from real problems \cite{etzioni_no_2016,crawfordcalo2016}, while some in the `long-term' camp suggest that the problems they focus on dwarf any near-term concerns in importance \cite{tegmark2017,bostrom_superintelligence_2014}. This adversarial dynamic seems to be at the core of proposals to `reconcile' or `bridge' the two sides \cite{baum2018,cave_bridging_2019}.

\section{The problem with the near/long-term distinction}

`Near-term' and `long-term' are often used in ways that combine and conflate multiple different dimensions, including: when issues arise; what kinds of technological capabilities issues relate to; how well-understood or speculative issues are; how high-stakes issues are and how many people they affect; and how long-lasting the consequences of an issue are likely to be. As table \ref{table1} shows, even a single definition or paper sometimes associates `near-term' with one of these dimensions and `long-term' with another: Brundage \cite{brundage_guide_2017}, for example, defines `near-term' based on when issues arise and `long-term' in terms of technological capabilities. 

Rather than describing a single dimension, we suggest that the distinction between near- and long-term as currently used is better understood as describing clusters of issues, incorporating multiple dimensions:

\begin{itemize}
    \item[] Issues considered \textbf{`near-term'} tend to be those arising in the present/near future as a result of current/foreseeable AI systems and capabilities, on varying levels of scale/severity, which mostly have immediate consequences for people and society.
    \item[]Issues considered \textbf{`long-term'} tend to be those arising far into the future as a result of large advances in AI capabilities (with a particular focus on notions of transformative AI or AGI), and those that are likely to pose risks that are severe/large in scale with very long-term consequences.
\end{itemize}

We believe that carving up the AI E\&S research space according to these clusters is problematic in two key ways. First, important and nuanced information gets lost if we give too much prominence to only two clusters. Many beliefs, values and assumptions will vary within each of these clusters. For example, among those who believe we should prioritise risks from advanced AI systems, there seems to be substantial disagreement about why exactly those risks are most important to work on, and what work should be done today to prevent those risks \cite{sittler2019,ngo2019}. Similarly, there are many different reasons a group may choose to focus on issues arising from current applications of AI in society: because they believe we should prioritise helping people alive today, because they believe advanced AI systems are very far off or hard to predict, or because they believe today's problems are as high-stakes as anything we might face in the future. As we will discuss later, a binary distinction may have led the research community to neglect impacts which fall somewhere between `immediate' and `very long-term' \cite{parson_artificial_2019}. 

Second, giving too much prominence to the `near-term' and `long-term' clusters without examining underlying beliefs can lead researchers who identify with either side to end up misunderstanding each other. For example, Etzioni \cite{etzioni_no_2016} argues against the idea that superhuman AI could pose an existential threat to humanity (most prominently argued by Bostrom, \cite{bostrom_superintelligence_2014}) by drawing on survey data to suggest that experts do not believe superhuman AI systems will arrive any time soon. But as Dafoe and Russell \cite{dafoe_yes_2016} point out in response, Bostrom's argument does not actually rely on any assumptions about superintelligence being imminent. In practice, Etzioni and Bostrom's disagreement turns more on how important and tractable they think it is to work on the potential risks from superhuman AI today. The idea that there are two `camps' of AI E\&S researchers, with very opposing perspectives, may lead researchers like Etzioni and Bostrom to make unfounded assumptions about the others' perspective, and prevent them from engaging in more substantive and productive debate. 

At its core, the distinction between near and long-term in AI E\&S aims to capture the idea that different researchers and groups will sometimes have quite different priorities which guide what they work on, underpinned by some quite fundamental disagreements. We believe that acknowledging these differences is useful, but suggest that we need a clearer way to think and communicate about what they are. In the following section, we propose exactly that - some clearer ways to communicate about different research priorities and underlying disagreements.

\section{A clearer account of research priorities and disagreements}

\subsection{Unpacking the near/long-term distinction}

As commonly used, the terms `near-term' and `long-term' in fact appear to capture four different dimensions of differing priorities within the AI E\&S research community:

\begin{itemize}
    \item[] \textsc{Capabilities}: whether to focus on the impacts and challenges of current AI systems, or those relating to much more advanced AI systems
    \item[] \textsc{Impacts}: whether to focus mostly on the immediate impacts of AI for society, or whether to consider possible impacts much further into the future.
    \item[] \textsc{Certainty}: whether to focus on impacts and issues that are relatively certain and well-understood, or those that are more uncertain and speculative.
    \item[] \textsc{Extremity}: whether to focus on impacts at all scales, or to prioritise focusing on those that may be particularly large in scale.
\end{itemize}

None of these four dimensions are binary: one can choose research questions that focus on AI systems that are more or less advanced, exploring impacts on varying time horizons, with varying degrees of certainty and extremity. Table \ref{table1} shows how these four dimensions in definitions of `near-term' and `long-term' by different authors.

\begin{table*}[t]
\begin{center}
 \begin{tabularx}{\textwidth} { 
  | >{\centering\arraybackslash\hsize=3cm}X 
  | >{\centering\arraybackslash}X 
  | >{\centering\arraybackslash}X |
  }
 \hline
 \large{\textbf{ Dimension }} & \large{\textbf{ 'Near-term'}} & \large{\textbf{ 'Long-term'}}\\ [0.5ex] 
 \hline \hline
\textsc{\newline\newline\newline AI Capabilities} & 
``Attention should go to existing and near-term AI'' \citep[p.2]{baum2018} & ``Issues that either only arise at all or arise to a much greater extent when AI is much more advanced than it is today, and in particular if advances turn out to be rapid.'' \cite{brundage_guide_2017}\newline\newline ``Attention should go to the potential for radically transformative long-term AI'' \citep[p.2]{baum2018}\\ 
 \hline
\textsc{\newline\newline\newline Impacts} & ``Issues society is grappling with today.'' \cite{brundage_guide_2017} \newline \newline ``Extant or imminently anticipated AI applications that interact with existing legal, political, or social concerns'' \citep[p.5]{parson_artificial_2019}\newline \newline`` [...] problem X that we are facing today with AI'' \cite{krakovna_is_2018} 
 & ``Longer-term concerns and opportunities that are less certain'' \citep[p.5]{cave_bridging_2019}

 \\ 
 \hline
\textsc{Certainty} & 
``Immediate challenges involving fairly clear players and parameters'' \citep[p.5]{cave_bridging_2019}
 & ``Longer-term concerns and opportunities that are less certain'' \citep[p.5]{cave_bridging_2019}
 \\ 
 \hline
\textsc{Extremity} &  & ``The existential risks of extreme AI advances'' \citep[p.5]{parson_artificial_2019}
 \\ 
 \hline
\end{tabularx}
\caption{Dimensions underlying the 'near-term' and 'long-term' distinction in the literature.}
\label{table1}
\end{center}
\end{table*}

Of course, these dimensions are not entirely independent of one another: for example, if one wants to focus on particularly large-scale impacts of AI, one may have to be willing to consider more speculative questions. However, it is still useful to unpack these different dimensions, particularly because there are many possible views and research questions which are not easily captured by the near/long-term distinction as commonly used. 

Looking at these four dimensions, we can identify some differences in what they refer to. `Capabilities' and `impacts' both refer in some sense to time horizons; to whether research focuses on the present (either in terms of AI capabilities or impacts) or whether it looks further into the future. Both these dimensions therefore seem to capture different \textit{interpretations} of what it might mean to say an issue is `near-term' or `long-term'. `Certainty' and `extremity', by contrast, do not relate so explicitly to time horizons: instead they relate to the nature of impacts that different researchers might focus on, and our ability to predict and understand  these impacts. Although these dimensions do appear in how near and long-term are described, we suggest that they better characterised as capturing the kinds of \textit{motivations} researchers may have for choosing to work on different areas.


\subsection{Distinguishing capabilities and impacts}

`Near-term' and `long-term' are often modifiers applied to the state of technological \textit{capabilities} themselves, with near-term issues being those arising from the AI systems we have today \cite{parson_artificial_2019}, and long-term issues being those we might expect to arise from certain kinds of much more advanced AI systems \cite{brundage_guide_2017}. At other times, `near-term' and `long-term' refer rather to the \textit{impacts} of the technology: with near-term issues being the impacts of AI that society already faces or will face in the immediate future \cite{brundage_guide_2017}, and long-term issues being the impacts of AI that society might face very far into the future \cite{cave_bridging_2019}.

Of course the timescale of technological advances and their impacts on society will be related. However, conflating long-term capabilities and impacts may mean the research community neglect important questions about the potential long-term impacts of current AI systems and their applications in society. For example, in what ways could injustices perpetuated by increased use of current algorithmic systems have very long-lasting and irreversible consequences for inequality in society? What are the possible longer-term implications of how data-intensive AI applications are beginning to change norms around privacy and personal data? \cite{newellmarabelli2015}

Furthermore, it is important to recognise that both capabilities and impacts lie on a spectrum between near- and long-term. Understanding the potential `medium-term' impacts of AI on society - for example, exploring what different sectors and domains might look like in 5-10 years as a result of current trends in AI development and deployment - is likely to be important both for thinking about what we can do today to protect against future harms, and for preparing for longer-term impacts and scenarios. Similarly, there are many ways in which AI systems could become more advanced over the coming years, before reaching anything close to `superintelligence' or `AGI'. We need to consider what issues these intermediate advances might raise, and what kinds of intermediate advances in capabilities are mostly likely to be of harm or benefit for society.

\begin{figure*}[t]
    \centering
    \includegraphics[width=.9\textwidth]{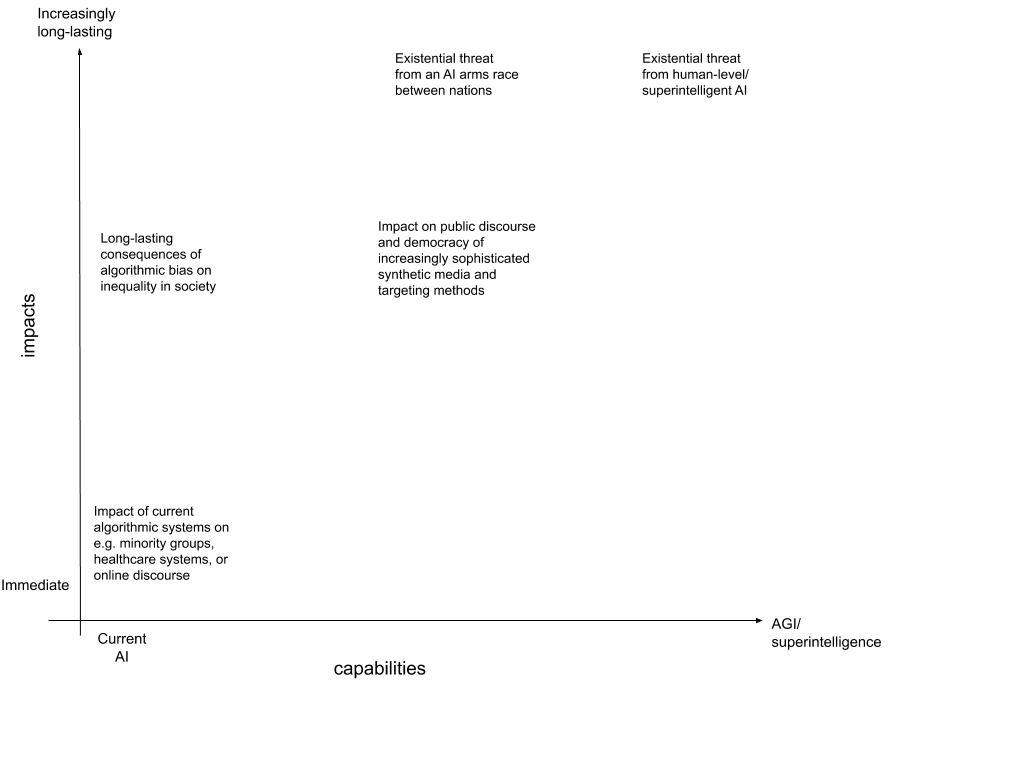}
    \caption{Distinguishing issues based on capabilities and impacts.} 
    \label{fig:1}
\end{figure*}

Figure \ref{fig:1} shows how considering these two dimensions of capabilities and impacts, each sitting on a spectrum, allows for more nuanced categorisation of different types of issues.

This way of conceptualising the research space could also be used to situate different research groups or research agendas. For example, AI Now's research seems to sit pretty squarely in the bottom left-hand corner \cite{whittaker_ai_2018}; the Centre for the Governance of AI at Oxford focuses mostly on issues at the top of the plot but spanning left to right \cite{dafoe_ai_2018}; and the Centre for the Future of Intelligence at Cambridge works on issues across a much wider range of the plot. Both authors of this paper see their research as mostly focusing in the upper-left quadrant - i.e. considering the medium-to-long-term impacts of near-to-medium-term AI capabilities, primarily because these issues have thus far been more neglected in discourse and research around AI E\&S impacts \cite{parson_artificial_2019}.

Of course, this still contains considerable ambiguity and room for disagreement, and placing different issues on these axes requires making assumptions about questions we don't necessarily have good answers to. Is there really good reason to think that advanced AI systems will pose an existential threat to humanity, or that advances in AI capabilities might pose a threat to democracy with long-lasting consequences? We believe that provoking productive debate and discussion on questions like these could in fact be a valuable consequence of this way of conceptualising the research space. However, we also recognise that any two-dimensional plot is limited in its ability to capture more fundamental reasons why people have those different priorities. We therefore next turn to a discussion of some of the beliefs and motivations underpinning different approaches in the AI E\&S research space.

\subsection{Underlying beliefs and motivations}

One question that has not been explored in existing discussions of the near/long-term distinction is \textit{why} different researchers and groups have different priorities. The idea of focusing on impacts that are more or less certain and more or less extreme are present in some definitions of near and long-term (see table \ref{table1}); we suggest that these dimensions may be best understood as capturing some of the underlying motivations people have for focusing on certain types of research questions.

One reason to focus on researching the longer-term impacts of AI, and/or the impacts of more advanced capabilities, is that those impacts might be more extreme and so particularly important to manage. Indeed, groups like the Open Philanthropy Project who explicitly prioritise risks from advanced AI state that ``all else equal, we're more interested in artificial developments that would affect more people and more deeply'' \cite{karnofsky_potential_2016}. Similarly, a research agenda published by the Future of Humanity Institute in Oxford, typically thought of as an institution with a `long-term' focus, explicitly highlights its ``focus on extreme risks'' \cite{dafoe_ai_2018}.

The idea that immediate impacts of AI and/or those relating to current AI systems are more certain and well-understood is also sometimes invoked as a reason to work on those issues (or as an argument against focusing on advanced capabilities and long-term impacts.) For example, Andrew Ng explains that the reason he is not concerned about advanced AI is ``the same reason I don't worry about overpopulation on Mars... we've never set foot on the planet so how can we productively worry about this problem now?'' \cite{garling_why_2015}. Relatedly, criticisms of those who focus on potential existential risks from AI often make the argument that these worries are too speculative, and are ``a distraction from the very real problems with artificial intelligence today'' \cite{crawford_artificial_2016}.

Working on problems that may have very extreme impacts on humanity, and working on problems that are certain and well-understood today are somewhat in tension with one another, and this tension is perhaps a central driver of the divide between `long-term' and `near-term' communities in AI E\&S. Of course, in line with the main message of this paper, our view is that there is not a simple binary choice between focusing on `extreme' impacts of AI and focusing on `certain' impacts - both exist on a spectrum, and there may be many ways for the AI E\&S community to identify areas for research which have relatively high-stakes implications while also being grounded enough in current AI and societal trends to make progress on. 

Further dimensions of disagreement may help explain why the AI E\&S community has divided in certain ways. Disagreement on \textit{normative} issues may be relevant here, such as around whether we have a special moral obligation to help those who are alive today over those who will live in future \cite{parfit_future_2017}, or to prioritise helping those worst off in society \cite{parfit_equality_1991}. Someone who holds the more fundamental philosophical belief that we should prioritise helping the worst off in society, for example, is likely to choose to work on the implications of AI for global fairness or social justice, regardless of their position on the certainty/extremity tension as outlined above. 

Other disagreements underpinning different priorities may be \textit{empirical}: disagreements about the nature of problems, or our ability to solve them. Whether one believes it is more important to focus on challenges relating to current or advanced AI systems is likely to depend at least in part on beliefs about AI progress. If you believe we are likely to see very fast and/or discontinuous progress in AI capabilities over the next few years, preparing for the potential impacts of these advances seems much more urgent than if you believe progress will be slow and gradual, giving us time to solve today's problems before we have to deal with anything more advanced. Indeed, much of the disagreement about whether we should be concerned about advanced AI seems to turn on disagreement about whether large advances in AI capabilities are possible any time soon \cite{brockman2019,ford2019}. Which specific impacts one chooses to focus on may also depend on empirical beliefs about the severity of different issues: how damaging to society threats to privacy might be in the long-term, and/or how likely advanced AI systems are to pose an existential threat, for example.

Beliefs about our ability to forecast and/or influence the future are also relevant - often scepticism about research on advanced AI systems seems to stem from doubts about our ability to productively work on these problems. Andrew Ng's statement quoted earlier \cite{garling_why_2015} appears to express this kind of scepticism: that we can't `productively' work on the impacts of advanced AI systems.

\section{Recommendations and conclusion}

The field of research concerned with the impact of AI on society and humanity is fast growing. Due to the cross-cutting and interdisciplinary nature of these issues, it is useful to be able to carve up the research space in ways that go beyond traditional disciplinary or thematic boundaries in academia, and capture different underlying assumptions about what the most important problems in this space are. We propose more nuanced ways to do this that better capture this rich and complex research space, breaking down the binary distinction between `near-term' and `long-term' into four different dimensions: capabilities, impacts, certainty and extremity, and emphasising that all of these sit on a spectrum.

Based on this analysis, we have a few concrete recommendations for how the AI E\&S research community can encourage more nuanced and productive discussion about different priorities and assumptions:

\begin{itemize}
    \item \textbf{Be specific about what you mean when using `near-term' or `long-term'} to refer to research priorities or projects. As outlined in the previous section, we think it would be particularly useful if there was a clearer distinction between (a) immediate vs. long-term impacts on society, and (b) current vs. much more advanced AI capabilities, and if there were greater acknowledgement that in both cases `near to long-term' is a spectrum, not a binary distinction.
    \item \textbf{Communicate clearly and explicitly about the assumptions and beliefs underlying your projects and priorities}, especially around some of the questions highlighted above. A good example of this is the preface to Dafoe's \cite{dafoe_ai_2018} research agenda, which explicitly states its focus on extreme risks from advanced AI (though they could go even further to explain the reasoning behind this decision.) Another good example is Parson et al. \cite{parson_artificial_2019} who clearly state that their project focuses on the ``intermediate scale of AI impacts, time horizons, and implications'' since this is where their expertise is most relevant, and where areas of potential importance are receiving relatively less attention.
    \item \textbf{Make an effort to understand the underlying motivations and assumptions of others with different research priorities}, again using some of the questions outlined in the previous section as a starting point. Conferences, workshops and journals could potentially help support this by providing fora for researchers to debate such fundamental questions and disagreements, improving mutual understanding.
\end{itemize}

By taking these steps, we hope the AI E\&S research community can: (a) establish new opportunities for collaboration and reduce adversarial dynamics, while allowing for differences of opinion and approach; (b) develop more consistent and coherent research agendas by helping researchers to think more clearly about what they prioritise; and (c) identify neglected research areas which may have been overlooked due to not fitting neatly into existing ways of characterising the research space.


\bibliographystyle{ACM-Reference-Format}
\balance
\bibliography{references}
\end{document}